\documentclass[twocolumn,showpacs,preprintnumbers,amsmath,amssymb]{revtex4-1}

\usepackage{graphicx}
\usepackage{dcolumn}
\usepackage{bm}
\usepackage{color}
\begin{document}

\preprint{APS/123-QED}

\title{Giant Anisotropic Magnetoresistance due to Purely Orbital Rearrangement  \\
in the Quadrupolar Heavy Fermion Superconductor PrV$_2$Al$_{20}$}

\author{Yasuyuki Shimura$^{1,2}$}
\email{simu@hiroshima-u.ac.jp}
\author{Qiu Zhang$^3$}
\author{Bin Zeng$^3$}
\author{Daniel Rhodes$^3$}
\author{Rico Uwe Sch$\ddot{\rm o}$nemann$^3$}
\author{Masaki Tsujimoto$^1$}
\author{Yosuke Matsumoto$^4$}
\author{Akito Sakai$^1$}
\author{Toshiro Sakakibara$^1$}
\author{Koji Araki$^5$}
\author{Wenkai Zheng$^3$}
\author{Qiong Zhou$^3$}
\author{Luis Balicas$^3$}
\author{Satoru Nakatsuji$^{1,6}$}
\email{satoru@issp.u-tokyo.ac.jp}

\affiliation{$^{1}$Institute for Solid State Physics, The University of Tokyo, Kashiwa, Chiba 277-8581, Japan \\
$^{2}$Graduate School of Advanced Sciences of Matter, Hiroshima University, Higashi-Hiroshima, 739-8530, Japan \\
$^{3}$National High Magnetic Field Laboratory, Florida State University, Tallahassee, Florida 32310, USA \\
$^{4}$Department of Quantum Materials, Max Planck Institute for Solid State Research, Heisenbergstrasse 1, Stuttgart 70569, Germany \\
$^{5}$Department of Applied Physics, National Defense Academy, Yokosuka, Kanagawa 239-8686, Japan \\
$^{6}$CREST, Japan Science and Technology Agency (JST), 4-1-8 Honcho Kawaguchi, Saitama 332-0012, Japan
}

\date{\today}

\begin{abstract}
We report the discovery of giant and anisotropic magnetoresistance due to the orbital rearrangement in a non-magnetic correlated metal.
In particular, we measured the magnetoresistance under fields up to 31.4 T in the cubic Pr-based heavy fermion superconductor PrV$_2$Al$_{20}$ with a non-magnetic $\Gamma _3$ doublet ground state, exhibiting antiferro-quadrupole ordering below 0.7 K.
For the [100] direction, we find that the high-field phase appears between 12 T and 25 T, accompanied by a large jump at 12 T in the magnetoresistance ($\Delta MR \sim $ 100 $\% $) and in the anisotropic magnetoresistivity (AMR) ratio by $\sim $ 20 $\% $.
These observations indicate that the strong hybridization between the conduction electrons and anisotropic quadrupole moments leads to the Fermi surface reconstruction upon crossing the field-induced antiferro-quadrupole (orbital) rearrangement.
\end{abstract}

\maketitle

Spintronic devices using both electronic charge and spin degrees of freedom have been developed,
 for instance, memory devices using the giant magnetoresistance (GMR) effect of ferromagnetic multilayers \cite{GMR_Baibich88}.
In addition to the spin and charge degrees of freedom, electronic orbital degrees of freedom have attracted much attention
 due to the discoveries of exotic orbital ordering and orbital liquid states \cite{Orbit_Tokura00, BCSO_Nakatsuji12}.
Moreover, since electronic orbitals coupling with lattice are responsible for forming the band structure, the orbital rearrangement should make dramatic effects on the transport phenomena.
Indeed, some perovskite-type manganese (Mn) oxides exhibit a gigantic negative magnetoresistance named as colossal magnetoresistance (CMR),
 induced by the suppression of the Mn orbital ordering under magnetic field \cite{CMR_Tokura06}.

On the other hand, to develop higher density memory device, it is important to find a mechanism for non-ferromagnetic materials to exhibit a large transport anomaly
 such as anisotropic magnetoresistance (AMR) and anomalous Hall effect without having spontaneous magnetization,
 as stray fields perturbing neighboring cells are absent \cite{Mn3Sn_Nakatsuji15, Mn3Ge_Kiyohara16, AFM_Jungwirth16}.
The AMR is defined as the difference between the resistances measured with currents applied parallel and perpendicular to the ordered spin direction, namely magnetic field. 
The AMR has been observed in the ferromagnetic and antiferromagnetic alloys \cite{AMR_McGuire75}.
In the case of antiferromagnets, the AMR effect has been limited to 1-2 $\%$ at room temperature \cite{MnTe_Kriegner16, FeRh_Marti14}.
Orbital ordering might be more useful for the observation of a large AMR not only because it should introduce anisotropy in the transport and in the electronic structure but because orbital moments are in principle non-magnetic.
In fact, strongly anisotropic transport has been reported near the putative quantum critical point of the orbital (nematic) ordering in the iron based superconductors \cite{FeAs_Shimojima10, FeAs_Chu10, FeAs_Kasahara12}.  

In 3$d$ transition metal compounds, however, the AMR may arise from various effects accompanied by the orbital ordering. 
First of all, the orbital ordering in the 3d systems is often induced by Jahn-Teller distortions, and thus the lattice distortion could lead to a large anisotropy in the resistance.
Furthermore, the orbital degrees of freedom cannot be decoupled from the spin degree of freedom, thus it is usually hard to neglect the magnetic contributions to the transport anisotropy \cite{MnO3_Kuwahara99}.

In sharp contrast, the strong spin-orbit coupling in $4f$ rare-earth materials may provide an ideal situation for the study of orbital physics. 
For example, purely-orbital effects can be studied in a cubic system that possesses non-Kramers rare-earth ions with even numbers of 4$f$ electrons such as Pr$^{3+}$.
Some of them exhibit a non-magnetic $\Gamma _3$ doublet ground state stabilized by the cubic crystalline-electric-field for the non-Kramers ions \cite{LLW}.
The $\Gamma _3$ doublet has only non-magnetic electric quadrupole $O^2_2$, $O^0_2$ and octupole $T_{xyz}$ moments, without magnetic dipole moments.
Normally, these degrees of freedom are lost at low temperatures by multipole ordering.
Moreover, compared to the transition metal systems, $4f$ electron systems have relatively low energy scales, and it is feasible to tune their ground states by external fields and pressure through quantum critical points \cite{QCP_Hilbert07, QCP_Gegenwart08, b-YbAlB4_Nakatsuji08}.

\begin{figure}[t]
	\begin{center}
		\includegraphics[width=80mm]{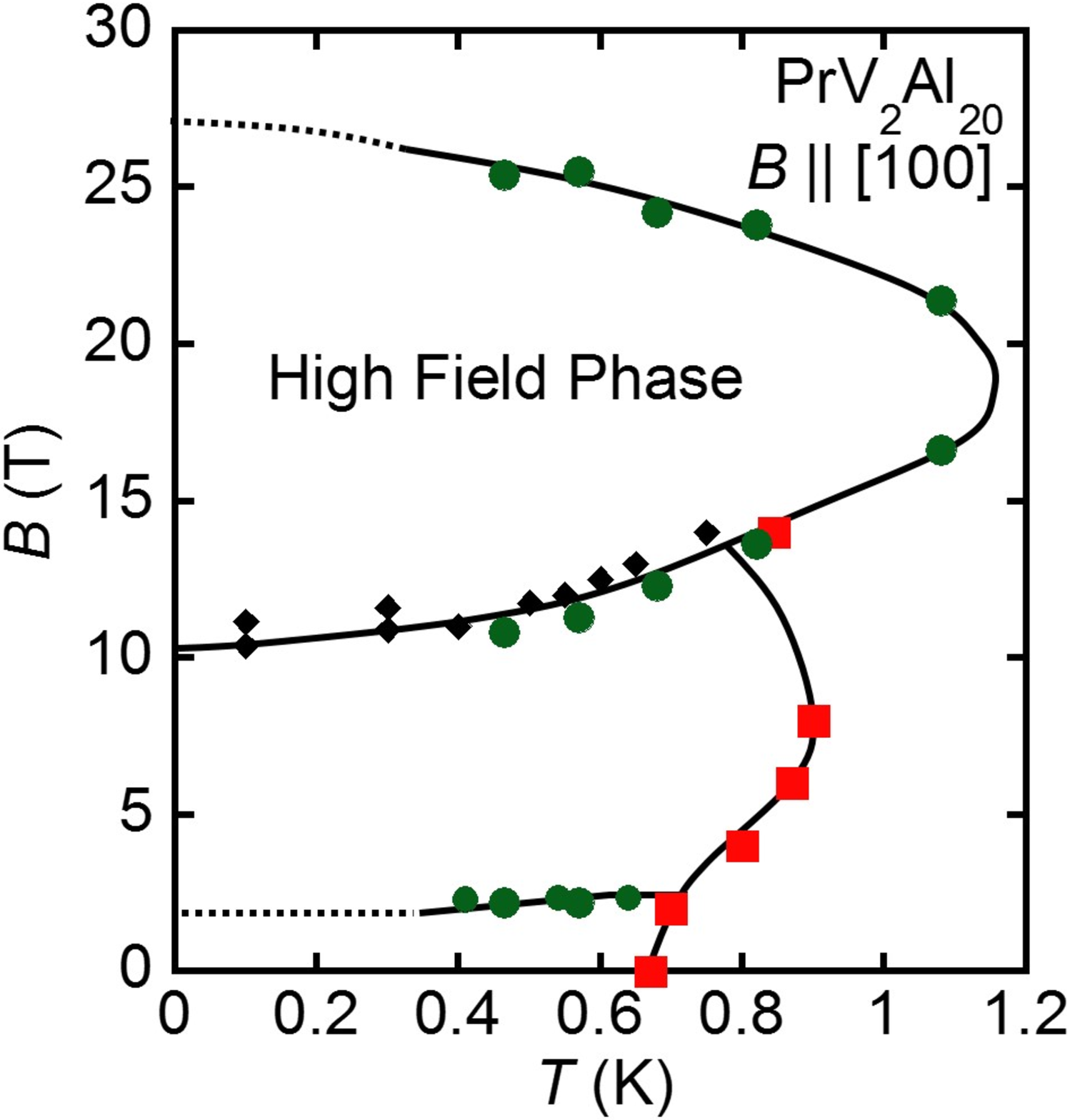}
	\end{center}
	\caption{(Color online) $B$-$T$ phase diagram for $B$ $||$ [100] obtained from the magnetoresistance $\rho (T, B)$.
		Circles indicate anomalies observed in $\rho (B)$, shown by open arrows in Fig. 2.
		Squares indicate the shoulder in $\rho (T)$ given in the supplemental material.
		Diamonds are plotted from the anomalies in the low-temperature magnetization $M(B, T)$ \cite{PrV2Al20_Shimura13}.
	}
\end{figure}

\begin{figure}[t]
	\begin{center}
		\includegraphics[width=80mm]{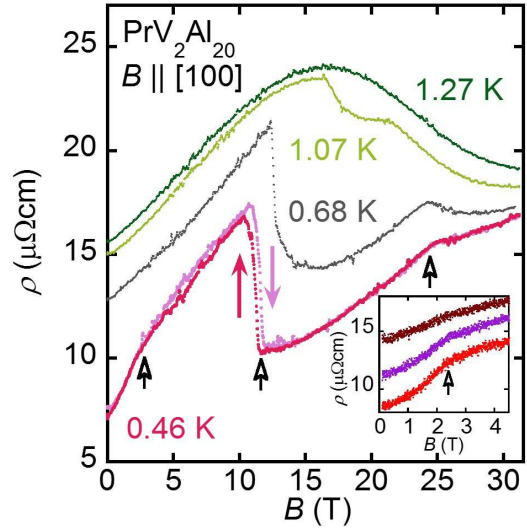}
	\end{center}
	\caption{(Color online) Field dependence of the magnetoresistance $\rho (B)$ for $B$ $||$ [100] below $\sim $ 1.3 K up to 31.4 T.
			The inset shows $\rho (B)$ below 4.5 T measured at 0.41 K (red), 0.54 K (purple), and 0.64 K (brown).
			The open arrows denote the transition fields. 
	}
\end{figure}

The cubic Pr-based compound PrV$_2$Al$_{20}$ has the $\Gamma_3$ ground doublet and exhibits antiferro-quadrupole ordering below 0.6-0.7 K \cite{PrTr2Al20_Sakai11}.
In the related compound PrTi$_2$Al$_{20}$, the $\Gamma_3$ ground doublet in the cubic crystalline electric field has been confirmed by inelastic neutron scattering measurements \cite{PrTi2Al20_Sato12}.
Within the quadrupolar ordered state, PrV$_2$Al$_{20}$ undergoes a transition into a heavy fermion superconducting phase below 0.05 K, demonstrating the strong hybridization between conduction electrons and the quadrupole moments \cite{PrV2Al20_Tsujimoto14}.
In addition, the strong hybridization makes the system proximate to a quadrupolar quantum critical point.
In fact, under magnetic fields along the [111] field direction, quantum critical behavior in the temperature dependence of the resistivity was observed around 11 T, where the quadrupole phase becomes fully suppressed \cite{PrV2Al20_Shimura15}.
On the other hand, field-induced exotic phenomena due to quadrupolar fluctuations are expected for other field directions.
Along the [100] field direction, another high-field phase was found above 11-12 T below $\sim $ 1 K $via$ low temperature magnetization measurements \cite{PrV2Al20_Shimura13}.
This high-field phase transition may well be induced by a rearrangement of quadrupole moments from the low-field antiferro order state.
Given the large hybridization, the anisotropic magnetoresistance across the quadrupolar transition is highly likely and it is thus quite interesting and important to study the magnetoresistance effects of PrV$_2$Al$_{20}$ under the field along [100]. 

In this paper, we report comprehensive results of the magnetoresistance measurements and the magnetic phase diagram of the quadrupolar ordered state in PrV$_2$Al$_{20}$ for the field parallel to the [100] direction.
Especially, we have discovered a sharp magnetoresistive jump accompanied by a large AMR
 through the field-induced transition at 12 T for $B$ $||$ [100]. 
The large and anisotropic magnetoresitance has never been reported for non-magnetic quadrupolar systems.
We attribute them to the reconstruction of the 4$f$ Fermi surface induced by the field-induced switching of the quadrupole orderings.
The experimental method is described in the supplemental information (SI) 

First as a summary, we present the magnetic phase diagram for $B$ $||$ [100] in Fig. 1,
 determined by the anomalies observed in the magnetoresistance $\rho (T, B)$.
The high-field phase was observed between $\sim $ 12 T and $\sim $ 24 T below $\sim $ 1.2 K.
This phase cannot be explained by the crossing of the crystalline-electric-field levels since the magnitude of the gap ($\sim 40$ K) between first-excited $\Gamma_5$ triplet and ground $\Gamma_3$ doublet is too large for such a level crossing \cite{PrV2Al20_Shimura13, PrV2Al20_Proc_Araki14, PrTr2Al20_Sakai11}.
For another cubic $\Gamma_3$ compound PrPb$_3$, Y. Sato $et$ $al$., has proposed a mechanism of the field-induced phase transition in the antiferro-quadrupole ordered system with $\Gamma _3$ ground doublet for $B$ $||$ [100] \cite{PrPb3_Sato10}.
Accordingly, an antiferro-quadrupole $O^2_2$ state is also expected to become stable under high $B$ $||$ [100] assisted by the octupole $T_\beta$ interaction.
As we show below and SI, the sharp anomalies seen in the field and temperature dependences of the resistivity across the phase boundaries into the high-field phase and the nearly absence of a corresponding change in the magnetization indicate that the high-field phase should be indeed an antiferro-quadrupole state. 

Figure 2 shows the field dependence of the magnetoresistance $\rho (B)$ for $B$ $||$ [100] ($\perp $ $I$ $||$ [011])
 below 1.3 K up to 31.4 T.
At 0.46 K, we observed three anomalies, a shoulder at 2 T, a distinct jump with a hysteresis at 12 T and a kink at 25 T.
As shown in the inset, a shoulder at $\sim $ 2 T does not exhibit clear temperature dependence up to $T_{\rm Q}$ $\sim $ 0.64 K.
The jump at 12 T is due to the phase transition by entering the high-field phase as also observed in the low-temperature magnetization measurements \cite{PrV2Al20_Shimura13}.
The hysteresis observed at 0.46 K indicates the first-order character of this transition.
With increasing temperature, two anomalies at 12 T and 25 T approach each other and disappear at 1.27 K.
In the SI, we show the temperature dependence of the resistivity $\rho (T)$ under fields up to 28 T for $B$ $||$ [100] 
In the temperature dependence, the kinks due to the transitions for low and high field phases are also detected. 
Significantly, the resistivity anomalies across the phase boundary into the high field phase are sharp. 
This clearly indicates that the high-field phase should not a ferro- but an antiferro-quadrupole order, as we discussed above.
From these anomalies in $\rho (T, B)$, the multiple phase diagram is constructed as shown in Fig. 1.
Above $\sim $ 25 T, we did not detect any anomalies in both $\rho (B)$ and $\rho (T)$, indicating that, for fields above 25 T, one stabilizes polarized para-quadrupole states.

\begin{figure}[t]
	\begin{center}
		\includegraphics[width=90mm]{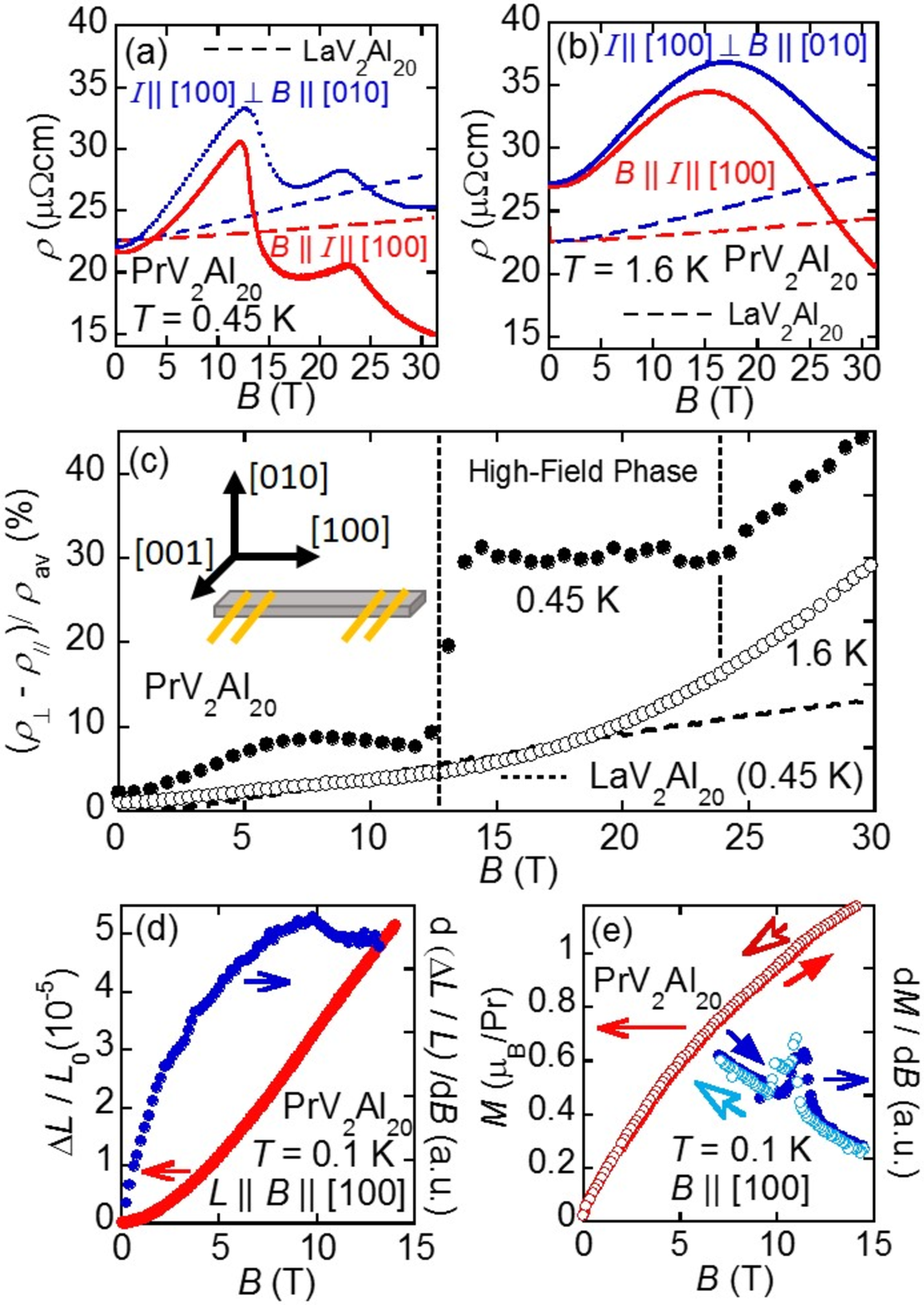}
	\end{center}
	\caption{(Color online) Longitudinal (a) and transverse (b) magnetoresistance
	 for $B$ $||$ $I$ $||$ [100] and $B$ $||$ [010] $\perp$ $I$ $||$ [100] at 0.45 K and 1.6 K up to 31 T in PrV$_2$Al$_{20}$, respectively.
	Dashed line indicates those in LaV$_2$Al$_{20}$.
	(c) displays the anisotropic magnetoresistance ratio $(\rho _\perp - \rho _{||})/\rho _{\rm av}$, where $\rho _{\rm av} = (2\rho _\perp + \rho _{||})/3$, obtained from the data in (a) and (b).
	(d) indicates the linear magnetostriction $\Delta L/L_0$ for $L$ $||$ $B$ $||$ [100] up to 14.5 T and d($\Delta L/L_0$)/d$B$ at 0.1 K.
	(e) displays the magnetization $M$ for $B$ $||$ [100] up to 14.5 T and d$M$/d$B$ at 0.1 K \cite{PrV2Al20_Shimura13}.
	}
\end{figure}

As shown in Fig. 2, we find a large magnetoresistance jump across the field-induced transition at 12 T for $B$ $||$ [100].
The magnitude of the change of the magnetoresistance at the transition is $\Delta MR = \Delta \rho _{\rm jump}/\rho _{0 \rm T} \sim 100 $ $\% $.
As we discussed, a similar high-field phase for $B$ $||$ [100] was also observed
 in the cubic antiferro-quadrupole ordered system PrPb$_3$ with the $\Gamma _3$ ground doublet \cite{PrPb3_Sato10}.
However, the magnetoresistance anomaly due to the field-induced phase transition in PrPb$_3$ is much smaller than our data for PrV$_2$Al$_{20}$ \cite{PrPb3_Yoshida16}.
The important character peculiar to PrV$_2$Al$_{20}$ is the strong hybridization between conduction electrons and quadrupole moments.
Indeed, the quadrupolar fluctuation is enhanced by the strong magnetic field as discussed for $B$ $||$ [111].
Thus, the field-induced quadrupole rearrangement most likely causes a reconstruction of the Fermi surfaces having a large 4f contribution through the strong hybridization, thus resulting in the large change in the magnetoresistance.

Since a quadrupole moment results from an anisotropic charge distribution,
 the band structure and the resultant transport properties should also become anisotropic by the quadrupole ordering.
In order to evaluate the anisotropy, we focus on the anisotropic magnetoresistance (AMR) ratio in the high-field phase above 12 T for $B$ $||$ [100].
AMR is defined as the difference between the longitudinal magnetoresistance $\rho _{||}$ and the transverse magnetoresistance $\rho _\perp $.
 
Figures 3 (a) and (b) show the magnetoresistance for two conditions of $B$ $||$ $I$ $||$ [100]
 (longitudinal, $\rho _{||}$) and $B$ $||$ [010] $\perp$ $I$ $||$ [100] (transverse, $\rho _\perp $) at 0.45 K and 1.6 K, respectively.
We also measured them in the reference compound LaV$_2$Al$_{20}$ without 4$f$ electrons.
Since these have a cubic crystal structure, the electronic state for $B$ $||$ [100] is intrinsically the same as that for $B$ $||$ [010].
Therefore, we may study the longitudinal and transverse magnetoresistance without changing the electronic state even under magnetic field.
Longitudinal and transverse conditions for the same sample were obtained by rotating the sample holder. 
These two field angles are precisely determined by the Hall sensors attached to the sample holder at low temperatures.
At 0.45 K inside the ordered phase in PrV$_2$Al$_{20}$, we observe two anomalies at 12 T and 23 T, which are almost consistent with those found in the magnetoresistance shown in Fig. 2.

Figure 3 (c) shows the AMR ratio $(\rho _\perp - \rho _{||})/\rho _{\rm av}$, where $\rho _{\rm av} = (2\rho _\perp + \rho _{||})/3$,  obtained from the data given in Figs. 3 (a) and (b).
In the high-field phase in PrV$_2$Al$_{20}$, the value of AMR is about 30 $\%$.
This large AMR is absent in the para-quadrupole state at 1.6 K. 
AMR in the reference compounds LaV$_2$Al$_{20}$ without 4$f$ electrons is only 5-10 $\%$ in this field region.
These indicate that 20-25 $\%$ AMR in the high-field phase in PrV$_2$Al$_{20}$ is purely due to the $4f$ contribution.
From these, the large AMR in the high-field phase in PrV$_2$Al$_{20}$ cannot be explained by the cyclotron motion inducing the transverse magnetoresistance.

To discuss the origin of the AMR in the high-field phase, now we examine the field dependence of the linear magnetostriction $\Delta L/L_0$ and the magnetization $M$ at 0.1 K up to 14.5 T for $B$ $||$ [100], where $L_0$ is the sample size at room temperature.
The magnetization data was already reported in Ref. \cite{PrV2Al20_Shimura13}.
The field derivatives d$\Delta L/L_0$/d$B$ and d$M$/d$B$ exhibit tiny but distinct anomalies at $\sim $ 12 T due to the crossing  of the phase boundary towards the high-field phase.
The clear jump in AMR ratio (20 - 25 $\% $) is very different from the field dependence of $\Delta L/L_0$ and $M$.
This weak anomaly in $M$ and $\Delta L/L_0$ suggests that the AMR in the high-field phase is not accompanied by any macroscopic lattice distortion and magnetization change.
Around 12 T, the magnitude of the tiny metamagnetic jump is just 0.6 $\% $ of 3.2 $\mu _{\rm B}$, the full moment of Pr$^{3+}$.
The value of $\Delta L/L_0$ at 12 T is just 0.005 $\% $, suggesting the cubic structure is almost preserved even in the high-field phase.
The AMR with almost no spontaneous magnetization change and distortion in the high-field phase probably comes from the non-magnetic field-induced antiferro-quadrupole ordering strongly coupling with the conduction electrons.
In comparison, the AMR ratio in the antiferromagnetic materials has been limited to 1-2 $\% $
 which is one order magnitude smaller than our observations \cite{FeRh_Marti14, MnTe_Kriegner16}.

Finally, we compare our results for PrV$_2$Al$_{20}$ with the case of iron-based superconductors Ba(Fe$_{1-x}$Co$_x$)$_2$As$_2$, which also exhibit a strongly anisotropic resistance almost without macroscopic distortion and spontaneous magnetization.
In BaFe$_2$As$_2$, the antiferromagnetic ordering almost coincides with a tetragonal to orthorhombic structural transition at 130 K \cite{FeAs_Huang08}.
By doping Co, these transition temperatures are strongly suppressed and disappear at the Co composition of $x \sim 0.07$, where superconducting transition temperature peaks \cite{FeAs_Nandi10}.
Significantly, in the low-temperature antiferromagnetic orthorhombic phase, the laser angle-resolved photoemission spectroscopy has revealed that the Fermi surface is mainly composed of iron $3d$ bands \cite{FeAs_Shimojima10}.
In the orthorhombic phase of BaFe$_2$As$_2$, the lattice parameters are a = 5.61587 $\text{\AA }$, b = 5.57125 $\text{\AA }$
 and c = 12.9428 $\text{\AA }$, almost preserving the tetragonal structure.
Note that, at $x \sim 0.04$ in the vicinity of the composition ($x = 0.07$) where the structural transition disappears,
 the ab-plane anisotropy in the resistance strongly develops up to $\rho _{I || \rm b}/\rho _{I || \rm a} \sim 1.8$,
 where the anisotropic resistance ratio is
 $(\rho _{I || \rm b}-\rho _{I || \rm a})/\rho _{\rm ave} \sim 60 $ $\% $ ($\rho _{\rm ave} = (\rho _{I || \rm b}+\rho _{I || \rm a})/2$) \cite{FeAs_Chu10}.
Above the temperatures of structural transition and superconducting dome, an electronic nematic state with a local $ab$-plane anisotropy was revealed by angle-resolved magnetic torque and synchrotron X-ray measurements \cite{FeAs_Kasahara12}.
While there would be still some effects due to $3d$ spin correlations, these observation suggest the $ab$-plane anisotropic resistance would originate from the electronic nematic order due to the $3d$ orbitals and not from a macroscopic structural distortion.

Very interestingly, a possible electronic nematicity has been also pointed out
 for the $4f$-electron based tetragonal antiferromagnet CeRhIn$_5$ \cite{CeRhIn5_Ronning17}.
Namely, a substantial anisotropy in the $ab$-plane magnetoresistivity was observed in the vicinity of a field-induced antiferromagnetic quantum critical point, suggesting the emergence of the electronic nematic state \cite{CeRhIn5_Ronning17}.
However, in this case as well, it is hard to isolate the pure orbital contribution from the magnetoresistance since the Ce ion has the $4f$-moment due to the Kramers degeneracy.

Here we note that our discovery in this paper provides a much clearer case without involving any spin degrees of freedom.
Namely, our observation of the large AMR in the high-field phase of PrV$_2$Al$_{20}$ for $B$ $||$ [100] results from the spontaneous change in the $c$-$f$ hybridized band structure induced by the rearrangement in the 4$f$-nonmagnetic orbitals (quadrupoles).
This orbital rearrangement causes almost no change in the magnetization and structure.


In conclusion, we have measured the magnetoresistance in the cubic antiferro-quadrupole ordered state of the heavy fermion superconductor PrV$_2$Al$_{20}$ with the non-magnetic quadrupolar doublet ground state and established the magnetic phase diagram for $B$ $||$ [100].
Upon entering the high-field phase at 12 T,
 we have discovered a large magnetoresistance jump with the magnetoresistance ratio 100 $\% $ accompanied by the large change in the anisotropic magnetoresistance by $\sim $ 20 $\% $.
These large changes in both magnetoresistance and anisotropic one are the consequences of the reconstruction of the Fermi surface
 at the field-induced quadrupole rearrangement due to strong hybridization between the conduction electrons and nonmagnetic 4$f$ quadrupole moments.

\begin{acknowledgments}
We thank K. Hattori, K. Matsubayashi, J. Suzuki and Y. Uwatoko for useful discussions.
This work is partially supported by CREST (JPMJCR15Q5), Japan Science and Technology Agency, Grants-in-Aid for Scientic Research (No. 25707030, 15J08663 and 25887015), by Grants-in-Aids for Scientific Research on Innovative Areas (15H05882, 15H05883) and Program for Advancing Strategic International Networks to Accelerate the Circulation of Talented Researchers (No.R2604) from the Japanese Society for the Promotion of Science. Y.S. is partially supported by the Institute of Complex Adaptive Matter (ICAM).
L.B. is supported by DOE-BES through award DE-SC0002613. The NHMFL is supported by NSF through NSF-DMR-1157490 and the State of Florida. 
\end{acknowledgments}
\end{document}